# Statistically equivalent models with different causal structures: An example from physics identity

Yangqiuting Li [1] and Chandralekha Singh [2]

[1]*Department of Physics, Oregon State University, Corvallis, Oregon 97331, USA*
[2]*Department of Physics and Astronomy University of Pittsburgh, Pittsburgh PA 15260, USA*

**Abstract.** Structural equation modeling (SEM) is a statistical method widely used in educational research to investigate relationships between variables. SEM models are typically constructed based on theoretical foundations and assessed through fit indices. However, a well-fitting SEM model alone is not sufficient to verify the causal inferences underlying the proposed model, as there are statistically equivalent models with distinct causal structures that equally well fit the data. Therefore, it is crucial for researchers using SEM to consider statistically equivalent models and to clarify why the proposed model is more accurate than the equivalent ones. However, many SEM studies did not explicitly address this important step, and no prior study in physics education research has delved into potential methods for distinguishing statistically equivalent models with differing causal structures. In this study, we use physics identity model as an example to discuss the importance of considering statistically equivalent models and how other data can help to distinguish them. Previous research has identified three dimensions of physics identity: perceived recognition, self-efficacy, and interest. However, the relationships between these dimensions have not been thoroughly understood. In this paper, we specify a model with perceived recognition predicting self-efficacy and interest, which is inspired by individual interviews with students in physics courses to make physics learning environments equitable and inclusive. We test our model with fit indices and discuss its statistically equivalent models with different causal inferences among perceived recognition, self-efficacy, and interest. We then discuss potential experiments that could further empirically test the causal inferences underlying the models, aiding the refinement to a more accurate causal model for guiding educational improvements.

## I. INTRODUCTION

Structural Equation Modeling (SEM) is a widely used statistical method for analyzing predictive relationships among variables [1]. It enables researchers to examine both the measurement properties of latent variables and the structural relationships between them [1]. SEM has proven to be highly valuable across diverse fields [1]. For instance, in physics education research, SEM can help researchers investigate predictive relationships between various factors and students' learning outcomes [2,3].

Conducting SEM analyses involves several steps, including model specification, model identification assessment, data collection, and model estimation and evaluation [1]. Model specification entails creating an SEM model grounded in prior theoretical studies. Model identification assessment determines the mathematical solvability of the model. Model estimation and evaluation entail testing the model using fit indices to gauge data fit. It is important to note that while a well-fitted SEM model can assist researchers in testing predictive relationships among studied variables, it is insufficient for establishing or verifying underlying causal relationships between the variables [4,5]. Previous research has indicated that in a predictive model [4,6], as long as independent variables are associated with dependent variables, they can be used as predictors. However, if researchers seek to evaluate the causal impact of independent variables on dependent variables, the accuracy of the model as a causal estimate depends on whether the model reflects the actual causal relationships between the variables [4,7]. In other words, predictors do not necessarily have a causal impact on the dependent variables. Therefore, relying solely on fit indices is not sufficient to verify the causal inferences underlying a proposed model since there exist statistically equivalent models, which fit the data equally well but have different causal structures. Therefore, one crucial element of the model estimation and evaluation step is considering statistically equivalent models and clarifying why the proposed model should not be rejected in favor of these alternatives [1]. However, this element is often overlooked in SEM studies, potentially undermining the robustness of research findings [8].

In this study, we use physics identity model as an example to demonstrate how an SEM model can have multiple statistically equivalent models with different causal structures, how one can specify a model based upon additional data and how the model can be refined further to converge on an even more causally accurate model. In particular, we specify a model based on our interviews with students to make physics learning environments equitable and inclusive as well as other supporting evidence. Then, we examine different equivalent models with their distinct causal structures. Finally, we discuss several experimental studies that could help further distinguish the equivalent models and determine a more causally accurate model.

## II. BACKGROUND AND GOAL

### A. General approach of SEM

Structural Equation Modeling (SEM) typically encompasses several key steps to analyze predictive relationships among variables. The first step is model specification, in which researchers establish the hypothesized relationships between observed variables and their corresponding latent variables, as well as the relationships among the latent variables [1]. This step relies on additional data, previous research and domain knowledge [1], providing the foundation for the subsequent steps of the analysis.

After specifying the model, researchers evaluate its identification [1]. Model identification refers to determining whether the model is under-identified, just-identified, or over-identified [1]. Model identification involves determining whether the model is under-identified, just-identified, or over-identified [1]. An under-identified model is characterized by having more parameters to estimate than available data points (variances and covariances of the observed variables) within the model. In contrast, a just-identified model possesses an equal number of parameters for estimation and data points, and an over-identified model holds more data points than parameters for estimation. To enable parameter estimation, a model must be either just-identified or over-identified [1]. Over-identified models carry particular significance in SEM as they enable researchers to assess model fit indices, which evaluate how effectively the tested model portrays the observed data [9].

Subsequently, researchers proceed to data collection, which involves gathering, preparing, and screening the data to ensure its quality and suitability for analysis [1]. Appropriate data selection and preparation are essential for obtaining reliable and valid results. Once the data is prepared, researchers estimate and evaluate the model using statistical software. The first step in estimation involves assessing how well the model fits the data [1]. Fit indices such as the chi-square test, Comparative Fit Index (CFI), and Root Mean Square Error of Approximation (RMSEA), are used to assess how well the model aligns with the observed data. If the fit is unsatisfactory, researchers may need to revise the model to improve its alignment [1].

In many previous studies involving SEM, researchers often proceed to interpret the parameter estimates and draw conclusions based on their model when the model fits the data well without explicitly clarifying the details of the model specification from among many statistically equivalent models. However, not clarifying the process of model specification and solely discussing good fit indices is inadequate for verifying the causal inferences underlying the proposed model, as there often exist statistically equivalent models of the proposed model that fit the data equally well but possess different causal structures [1]. Statistically equivalent models are a collection of models that yield identical correlation matrices, fit functions, chi-square values, and goodness-of-fit indices [10]. Consequently, these equivalent models can equally well explain the data compared to researchers' preferred model but might lead to different causal claims [1]. Therefore, researchers should explicitly acknowledge the existence of equivalent models, and provide reasons for favoring their preferred model over the equivalent versions [1]. Extra evidence supporting the specified model is necessary in this regard. The more robust the supporting evidence is to favor the specified model, the more causally accurate that model would be.

### B. Randomized experiment as a method for establishing causal inference

Randomized experiments, involving the random assignment of subjects into treatment and control groups, are often considered the "gold standard" for establishing causal inferences due to their ability to balance potential confounding factors on average [11]. Randomized experiments typically incorporate several design elements that enhance internal validity [12]. Firstly, the manipulation of the independent variable occurs prior to the measurement of the outcome (dependent variable). Secondly, the control group functions as a counterfactual benchmark for the experimental (treatment) group. Thirdly, randomization ensures that the independent variable is uncorrelated with other potential causes of the outcome.

In addition to establishing causal relationships between two variables, the exploration of mediation relationships among multiple variables offers valuable insights to researchers. Mediation involves the transfer of causality from an independent variable to a dependent variable through a third variable known as a "mediator." Previous studies recommend an approach to testing mediation hypotheses [13-16]. This involves conducting two distinct experiments: one that manipulates the independent variable and another that manipulates the hypothesized mediator [13]. Moreover, when a collection of experiments manipulates the independent variable and another set manipulates the mediator, synthesizing these two sets of experiments through meta-analysis can yield even more robust evidence for mediation [13]. In this paper, we will discuss how experimental studies can be used to test different statistically equivalent models with regard to identifying the ones that are more causally accurate based upon how consistent they are with the results of experimental interventions.

### C. Goal

In this study, we investigate the issue of SEM model equivalence in the context of physics identity framework. We begin by specifying a model based on our individual interviews with students in physics courses to make physics learning environments equitable and inclusive as well as other supporting evidence from prior studies. Then, we examine different equivalent models with their distinct causal structures. Furthermore, we suggest a range of experimental studies that could further help distinguish among these equivalent models to determine a more causally accurate model.

## III. THEORETICAL FRAMEWORK

### A. Physics identity framework

Prior studies have shown that physics identity is a crucial motivational factor for explaining students' participation in physics related careers [3,17]. Physics identity pertains to students' perception of themselves as "physics people" and influences their career decisions and academic goals [17]. Prior studies have identified three interrelated dimensions of physics identity: perceived recognition by others as a physics person, physics self-efficacy and interest [17]. These dimensions have been shown to be important predictors of students' overall physics identity [3,18].

Perceived recognition in a domain, such as physics, refers to students' perception about whether other people see them as a physics person [19]. Prior studies have shown that perceived recognition is the strongest predictor of students' overall physics identity compared to self-efficacy and interest [3,20,21]. Moreover, perceived recognition also predicts students' course grades in introductory physics courses [22,23].

Self-efficacy, defined as students' beliefs in their capability to succeed in a certain situation, task, or particular domain [20,24,25], can influence students' engagement and performance in a given domain [26,27]. Students with high self-efficacy often enroll in more challenging courses than those with low self-efficacy because they perceive difficult tasks as challenges rather than threats [28].

Interest is defined by positive emotions accompanied by curiosity and engagement in particular content [29]. Interest has also been shown to influence students' learning outcomes [26,29,30]. For example, one study showed that making science courses more relevant to students' lives and transforming curricula to promote interest in learning can improve students' achievement [31].

### B. Relationships between perceived recognition, self-efficacy, and interest

Research suggests that perceived recognition, self-efficacy, and interest correlate to and interact with each other [24,32], but the predictive relationships among them are not very clear. Prior studies have proposed different models with certain relationships among them. For example, some prior studies used a model in which self-efficacy is the predictor of both interest and perceived recognition [3,33], while another study used the model in which interest is the predictor of both self-efficacy and perceived recognition [34]. Although most of these studies have presented theoretical frameworks for their proposed models, they have not explicitly discussed the existence of statistically equivalent models and how their proposed models are favorable compared to the equivalent ones based on evidence beyond model fit indices [35].

In this study, we specify an SEM model in which perceived recognition predicts self-efficacy and interest, and self-efficacy predicts interest. The schematic representation of the SEM model is shown in Figure 1. This model draws inspiration from our previous qualitative research involving individual interviews with students in physics courses to make physics learning environments equitable and inclusive [36-40] as well as findings from prior studies [24,32].

In particular, the paths from perceived recognition to self-efficacy and interest in the model (Figure 1) draw inspiration from our prior hour-long individual interviews with 70 undergraduate student volunteers (55 women and 15 men including some unpublished data) in physics courses about their experiences in physics learning environments using a semi-structured protocol [22,36,37,39,41]. Our interview data show that women were less likely than men to feel positively recognized by physics instructors/TAs, and this lack of recognition or discouraging feedback from instructors/TAs deteriorated their self-efficacy as well as interest, and lowered self-efficacy further lowered their interest [22,36,37,39,41]. For example, some interviewed women reported that when they went to the course instructor or TA to ask for help on physics problems, sometimes they were explicitly told that the problems were "easy", "obvious" or "trivial", which they perceived as disparaging or belittling (negative perceived recognition) in that they felt they were being told that they are not smart enough to do physics if they could not do such easy problems on their own [22,36,37,39,41]. In addition, many interviewed women noted that their instructors/TAs sometimes showed more interest in male students' questions and answered male students' questions with more attention than when they answered their questions (negative perceived recognition) [36,41]. Moreover, some interviewed women reported that men in their physics courses were generally praised more by the instructors/TAs than women, and sometimes instructors/TAs called men who answered the questions "brilliant", which made them feel as though they were not brilliant [22,36]. In the interviews, women with these types of negative perceived recognition reported that these experiences affected their physics self-efficacy and interested. In particular, because of the negative experiences in their physics courses, they started questioning "Why am I here in the first place? Am I really interested in this?", and some confided that they had contemplated switching out of their major (either engineering or physics) while men never expressed similar concerns [36]. Some female students noted that these negative experiences made them wonder whether they were experiencing them because their questions were not good or too easy, and thus they started doubting their ability to do well in this course [36,41].

On the other hand, our interview data show that positive encouragement and recognition are likely to boost students' self-efficacy and interest in physics [22,36,37,39,41]. For example, some students reported that they feel better when the instructors know what students tend to struggle with and acknowledge that it's okay to not completely understand the content and they just need more practice, and eventually they will get over the struggle with that topic and move on to the next topic [36]. In addition, some interviewed students also mentioned that they felt really encouraged when their physics problem solving, posters and talks were recognized by their instructors [36]. Thus, our interview data show that students' perceived recognition from instructors/TAs plays an important role in shaping students' physics self-efficacy and interest [36].

In addition, the path from perceived recognition to self-efficacy also draws inspiration from Bandura's social cognitive theory [24], which suggests that individual's self-efficacy can be influenced by social persuasion (i.e., encouragement and discouragement pertaining to their performance or ability to perform). For example, in the educational context, prior studies have shown that constructive feedback and recognition from instructors can significantly enhance students' self-efficacy in writing [42,43]. Similarly, another study showed that personalized messages of encouragement provided by instructors on students' work can elevate their self-efficacy [44].

The path from perceived recognition to interest is also inspired by findings from other prior studies [36-40,45-47]. For example, Shanab and colleagues [46] found that positive verbal feedback during a puzzle-solving task prompted undergraduates to invest more time and rate their interest higher, compared to participants in a neutral feedback control group. Similarly, a meta-analysis conducted by Deci et al. demonstrated that verbal recognition positively influenced self-reported interest in both children and college students [47].

The path from self-efficacy to interest is also guided by previous research. For instance, vocational investigations have shown that self-efficacy in a domain can foster interest in activities [48-50]. Silvia delved into this issue through the lens of emotion psychology, revealing that one's self-efficacy in a domain can influence their perceived uncertainty about an activity's outcome, which in turn affects their interest [51-53].

As shown in Figure 1, there are paths from gender to perceived recognition, self-efficacy, and interest, and physics identity. These paths are inspired by prior studies showing gender differences disadvantaging women in these constructs [2,3,17,18,54,55]. For instance, one study reveals that women who receive A grades reported similar levels of self-efficacy as men who earn C grades at the end of a two-semester calculus-based introductory physics course [55]. Prior studies have suggested that factors such as societal stereotypes and biases about who belongs in physics and who can excel in physics can contribute to the gender gaps [56-60]. Additionally, the results of moderation analysis provided support for conducting the SEM analysis involving gender (further details about moderation analysis will be discussed later in this paper).

In this study, we first evaluated the fit indices of the model we specified inspired by our interviews to make physics learning environment more inclusive and equitable as well as other supporting evidence discussed. Then, we examined different equivalent models with their distinct causal structures. Finally, we propose several experimental studies that could help further distinguish the equivalent models and determine a more causally accurate model.

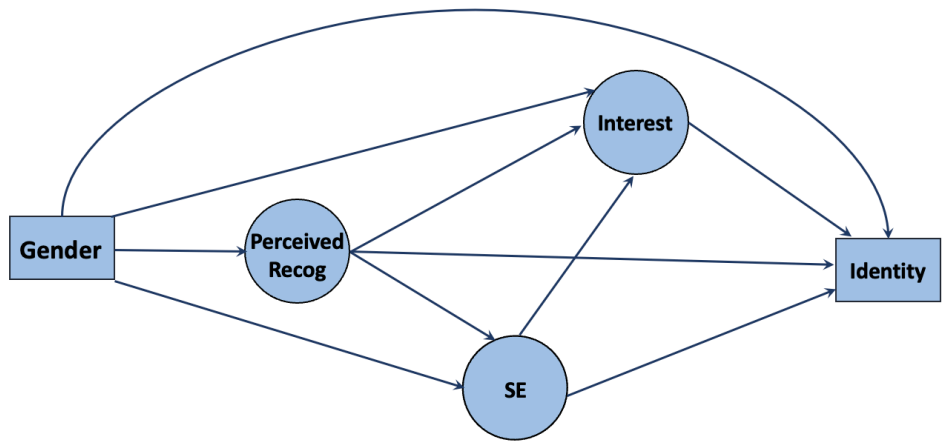

FIG. 1. Schematic representation of the SEM model we specified in which perceived recognition (Recog) predicts self-efficacy (SE) and interest, and self-efficacy predicts interest.

## IV.    RESEARCH QUESTIONS

Our research questions are as follows:

**RQ1.** Are there gender differences in students' self-efficacy, interest, perceived recognition, and overall physics identity at the end of the physics course studied?

**RQ2.** How well does the SEM model we specified fit the data?

**RQ3.** How many statistically equivalent models does our model have?

**RQ4.** How can one further determine a more accurate causal model using experimental studies?

## V.    METHODOLOGY

### A.  Participants

In this study, we collected motivational survey data at the end of the semester from students who took the introductory calculus-based physics 1 course in two consecutive fall semesters. This course is taken mostly by students majoring in engineering, physical sciences, and mathematics. The paper surveys were handed out and collected by TAs in the last recitation class of the semester. Finally, we combined the two semesters' data. The demographic data of students—such as gender—were provided by the university. Students' names and IDs were de-identified by an honest broker who provided each student with a unique new ID (which connected students' survey responses with their demographic information). Thus, researchers could analyze students' data without having access to students' identifying information.

There were 1219 students participating in this survey including both semesters. In our final data analysis, we kept 1203 students (including 427 women and 776 men) because the other 16 students did not provide their gender information. The data used in this study was also used in a prior publication for different research questions [2]. We recognize that gender is a social construct and is not binary. However, because students' gender information was collected by the university, which offered binary options, we did the analysis with the binary gender data in this study. 1.3% of the students who did not provide this information were not included in this analysis.

### B.  Survey instruments

In this study, we considered four motivational constructs—physics self-efficacy, interest, perceived recognition, and identity. The survey items for each construct are listed in Table I. The survey items were adapted from the existing motivational research [17,61-63] and have been re-validated in our prior work [64-66]. The validation and refinement of the survey involved use of one-on-one student interviews with both introductory and advanced students, exploratory and confirmatory factor analyses (EFA and CFA) [67], correlation between different constructs and Cronbach alpha (which is a measure of the internal consistency of each construct with several items) [68-70].

In our survey, each item was scored on a 4-point Likert scale (1-4). Students were given a score from 1 to 4 with higher scores indicating greater levels of interest, self-efficacy, perceived recognition and identity. Physics self-efficacy represents students' belief about whether they can excel in physics. We had four items for self-efficacy (Cronbach's alpha = 0.8) and these items had the response scale "NO!, no, yes, YES!". We also had four items for interest (Cronbach's alpha =

0.82). The question "I wonder about how physics works" had temporal response options "Never, Once a month, Once a week, Every day", whereas the question "In general, I find physics" had response options "very boring, boring, interesting, very interesting". The remaining two items were answered on the "NO!, no, yes, YES!" scale. Physics identity corresponds to students' belief about whether they designate themselves as a physics person [17]. Perceived recognition corresponds to whether a student thinks other people see them as a physics person [17,71,72], and it includes three items which correspond to family, friends and TA/instructor (Cronbach's alpha = 0.86). These items involved a four-point Likert response on the scale: "strongly disagree, disagree, agree, and strongly agree" and they correspond to 1 to 4 points [73].

**Table I.** Survey items for each of the motivational constructs, along with CFA factor loadings. Lambda represents the factor loading of each item, which is the correlation between the item and the construct with $p < 0.001$ indicating the correlation is highly statistically significant. The square of Lambda for each item gives the fraction of its variance explained by the construct. †The response options for this question are "Never, Once a month, Once a week, Every day". ‡The response options for this question are "very boring, boring, interesting, very interesting".

| Construct and Item | Lambda | p value |
|---|---|---|
| **Physics Identity** | | |
| I see myself as a physics person. | 1.000 | <0.001 |
| **Physics Self-Efficacy** (Cronbach's alpha = 0.8) | | |
| I am able to help my classmates with physics in the laboratory or in recitation. | 0.796 | <0.001 |
| I understand concepts I have studied in physics. | 0.829 | <0.001 |
| If I study, I will do well on a physics test. | 0.787 | <0.001 |
| If I encounter a setback in a physics exam, I can overcome it. | 0.742 | <0.001 |
| **Physics Interest** (Cronbach's alpha = 0.82) | | |
| I wonder about how physics works † | 0.710 | <0.001 |
| In general, I find physics ‡ | 0.893 | <0.001 |
| I want to know everything I can about physics. | 0.854 | <0.001 |
| I am curious about recent physics discoveries. | 0.748 | <0.001 |
| **Physics Perceived Recognition** (Cronbach's alpha = 0.86) | | |
| My family sees me as a physics person. | 0.925 | <0.001 |
| My friends see me as a physics person. | 0.940 | <0.001 |
| My physics TA and/or instructor sees me as a physics person. | 0.780 | <0.001 |

### C. Data analysis

#### 1. Descriptive statistics

In this study, we calculated the mean score for each construct for women and men. We note that all motivational constructs studied were measured using 4-point Likert scale survey items and each item is a categorical variable. In our previous study [20], we have checked the response option distances for our survey constructs by using item response theory (IRT) to support the use of means across ratings [74]. Here, we performed IRT with the new data set to verify the validity of using means across ratings. The parametric grades response model (GRM) by using the R software package "mirt" was used to test the measurement precision of our response scale [75,76]. Some of the items have response scales of "strongly disagree, disagree, agree, and strongly agree" while other items had response scale "NO!, no, yes, YES!". GRM calculates the location parameter for each response and calculates the difference between the locations. For the first group—strongly disagree, disagree, agree, and strongly agree—the differences between the location parameters were 1.3 and 1.4. For the second group — "NO!, no, yes, YES!" — the differences between the location parameters were 1.4 and 2.0. These results show that the numerical values for the location differences for item responses are comparable, which suggests that calculating the traditional mean score for items is reasonable [74,76]. Furthermore, we estimated the IRT-based scores with expected a posteriori (EAP) computation method for each construct, and the results are highly correlated with the mean scores (the correlation coefficient are > 0.98 for all constructs), which indicates that the use of mean scores is reasonable [74].

Before investigating the gender differences in the studied constructs, we assessed the normality of mean scores for items under each construct using skewness and kurtosis. Bulmer suggests that skewness values between -0.5 and 0.5 characterize a symmetric distribution, while values of -1.0 to -0.5 and 0.5 to 1.0 indicate a moderate degree of skewness, and values less than -1.0 and greater than 1.0 represent a high degree of skewness [77]. Other literature suggests that data is considered normal if skewness falls between -2 and +2 and kurtosis between -7 and +7 [78,79]. As shown in Table II, while the skewness of interest suggests a moderate degree of skewness based on the strict criteria mentioned earlier, most values of skewness and kurtosis shown fall within the normal range. Since physics identity is a categorical variable measured by a single item, we did not calculate Skewness and Kurtosis for it. In this study, we used the Wilcoxon rank-sum test to estimate the gender differences in the constructs studied. The Wilcoxon rank-sum test is commonly used to compare two independent samples when normality assumption is not satisfied or the data are ordinal [80].

**Table II.** Summary of the skewness and kurtosis

| Constructs | Skewness | | Kurtosis | |
|---|---|---|---|---|
| | Statistic | Std. Error | Statistic | Std. Error |
| Self-efficacy | -0.50 | 0.07 | 0.79 | 0.14 |
| Interest | -0.58 | 0.07 | 0.48 | 0.14 |
| Perceived Recognition | -0.13 | 0.07 | -0.44 | 0.14 |

### *2. Structural Equation Modeling*

In this study, we used the R [81] software package "lavaan" to conduct Structural Equation Modeling (SEM) [82] to investigate the relationship between students' perceived recognition, interest and self-efficacy. SEM is a multivariate statistical analysis technique that is used to model the relations between observed variables (items) and latent variables (factors), or between multiple latent variables. This technique is the combination of confirmatory factor analysis (which tests how well the observed variables represent the latent variables) and path analysis (which estimates the regression relationships between latent variables). Compared with a multiple regression model, a major advantage of SEM is that we can estimate all of the regression links for multiple outcomes and factor loadings for items simultaneously, which improves the statistical power [82]. Another advantage of SEM is that it shows not only the direct regression relation between two constructs but also all the indirect relations mediated through other constructs [82].

In SEM, it is generally recommended that ordinal Likert-type items with more than 5 categories can potentially be treated as continuous indicators eligible for the Maximum Likelihood (ML) estimator [83]. In our case, where we employed 4-point Likert scale, the use of polychromic correlations is recommended. In addition, in this study, we used diagonally weighted least square (DWLS) to estimate parameters. DWLS estimation is commonly used to analyze ordinal variables and has also been shown to produce unbiased parameters estimates with great statistical power for ordinal data [84,85].

As noted earlier, the SEM includes two parts: confirmatory factor analysis (CFA) and path analysis. First, we performed the CFA for each construct. The model fit is good if the fit parameters are above certain thresholds. In CFA, Comparative Fit Index (CFI) > 0.9, Tucker-Lewis Index (TLI) > 0.9, Root Mean Square Error of Approximation (RMSEA) < 0.08 and Standardized Root Mean Square Residual (SRMR) < 0.08 are considered acceptable and RMSEA < 0.06 and SRMR < 0.06 are considered a good fit [68]. In our study, CFI = 0.997, TLI = 0.996, RMSEA = 0.057 and SRMR = 0.040, which represents a good fit [68].

Before performing the path analysis, we calculated the pairwise correlations between each pair of constructs studied (see Table III) [70]. The correlation coefficients were calculated using the R software package "lavaan," employing the DWLS estimator, which is a common approach for estimating correlations between variables involving categorical data [86,87]. As shown in Table II, the correlation coefficients of all constructs are above 0.6, and most of them are below 0.8, which indicates that even though they have correlations with each other, the correlations are not so high that the constructs could not be separately examined in the SEM [88]. We note that the correlation coefficient between physics identity and perceived recognition is 0.89. This is consistent with Godwin et al. and Kalender et al.'s prior work showing that perceived recognition (external identity) is the strongest predictor of physics identity (internal identity) [3,20]. In addition to the correlations between the constructs studied, we report the correlations between all measured items in Appendix A.

**Table III**. Pairwise correlation coefficients of the constructs studied. *p* values are indicated by *** for $p < 0.001$.

| Constructs | 1 | 2 | 3 | 4 |
|---|---|---|---|---|
| 1. Physics identity | -- | -- | -- | -- |
| 2. Self-efficacy | 0.74*** | -- | -- | -- |
| 3. Interest | 0.75*** | 0.64*** | -- | -- |
| 4. Perceived Recognition | 0.89*** | 0.77*** | 0.70*** | -- |

Since the SEM model in this study involves gender, we conducted a moderation analysis [1,89] to test whether gender moderates the relationship between any two constructs in the model (i.e., do the strength of relationships given by the standardized regression coefficients between any two constructs in the model differ for women and men?). We used the R [81] software package

"lavaan" to conduct multi-group SEM. We initially tested for measurement invariance, which includes testing of factor loadings, indicator intercepts and residual variances. Then, we investigated whether the regression pathways were different across gender. Results showed that in all of our models, strong measurement invariance holds and there is no difference in any regression coefficients by gender, which allowed us to perform the path analysis involving gender using SEM [1,89] as shown schematically in Fig. 1.

We first analyzed the saturated SEM model that includes all possible links from left to right between different constructs shown in Figure 1, and then we removed the most insignificant path line (with the highest $p$ value) and re-ran the model. We used this method to trim one path at a time until all remaining path lines were statistically significant [89].

## VI. RESULTS

### A. Descriptive statistics

Pertaining to RQ1, Table IV shows the descriptive statistics of women's and men's physics identity, perceived recognition, self-efficacy, and interest, along with the results of Wilcoxon rank-sum tests for gender differences. Cohen suggested that typically values of 0.1, 0.3 and 0.5 represent small, medium and large effect sizes for Wilcoxon rank-sum tests [90]. As shown in Table IV, women have significantly lower average scores in all four motivational constructs [91]. In particular, women's average scores on physics identity and perceived recognition were below the neutral score of 2.5. Thus, many women did not think others see them as a physics person, and they did not see themselves as a physics person either. In Appendix B, we report the percentages of students who selected each choice for each survey item, which show consistent results with the descriptive statistics shown in Table IV.

**Table IV**. Descriptive statistics for women and men, in which M stands for construct mean value, SD is the standard deviation and N is the number of students. Effect sizes and $p$-values are presented in the right most column with $p < 0.001$ indicating highly statistically significant gender differences. A minus sign indicates that men have higher scores than women.

| Constructs | Women N = 427 | | Men N = 776 | | Statistics | |
|---|---|---|---|---|---|---|
| | M | SD | M | SD | Effect size | $p$ value |
| Physics Identity | 2.17 | 0.83 | 2.63 | 0.83 | -0.26 | < 0.001 |
| Perceived Recognition | 2.24 | 0.72 | 2.60 | 0.73 | -0.23 | < 0.001 |
| Self-efficacy | 2.71 | 0.57 | 2.99 | 0.50 | -0.23 | < 0.001 |
| Interest | 2.72 | 0.64 | 3.10 | 0.58 | -0.29 | < 0.001 |

### B. Estimation of the specified SEM model

Pertaining to RQ2, we estimate how the specified SEM model (Model 1) in Figure 1 fit the data. Fig. 2 shows the results of the SEM model. The solid lines represent regression paths, and numbers on the lines are regression coefficients ($\beta$ values), which represent the strength of regression relations. A regression coefficient reflects the change in the dependent variable

(outcome) associated with a one-standard-deviation increase in the independent variable (predictor), while holding other variables in the model constant [92]. A summary of all direct and indirect effects can be found in Table V.

The level of SEM model fit is represented by the Comparative Fit Index (CFI), Tucker-Lewis Index (TLI), Root Mean Square Error of Approximation (RMSEA) and Standardized Root Mean Square Residuals (SRMR), and CFI > 0.9, TLI > 0.9, RMSEA < 0.08, and SRMR < 0.08 are considered as acceptable (Hooper et al., 2007). The model in Fig. 2 fits the data well with CFI = 0.998, TLI = 0.998, RMSEA = 0.060 and SRMR = 0.044 [93]. Apart from evaluating the overall fit indices, we assessed the model's local fit. This involves examining the residual correlations among the studied items. The results of the local fit evaluation are detailed in Appendix C, confirming that our model fits the data well.

As shown in Figure 2, gender directly or indirectly predicts perceived recognition, interest, and self-efficacy, which is consistent with the descriptive statistics in Table IV, showing that women had statistically significantly lower score on these three constructs. Figure 2 also shows that perceived recognition, interest, and self-efficacy are all significant predictors of physics identity and perceived recognition is the strongest predictor (with $\beta = 0.52$), which is also consistent with prior studies by Godwin et al. [3] and Kalender et al. [20] showing that how students perceive themselves as a physics person is significantly influenced by their perception of how others view them as a physics person. Moreover, Figure 2 shows that perceived recognition also indirectly predicts physics identity through self-efficacy and interest. Therefore, the total effect of perceived recognition on physics identity is $\beta = 0.52 + 0.44 \times 0.31 + 0.71 \times 0.17 + 0.71 \times 0.27 \times 0.31 = 0.84$. We note that although there is a significant gender difference in physics identity (as shown in Table 3), gender does not directly predict physics identity, which indicates that the gender indirectly predicts physics identity through perceived recognition, interest, and self-efficacy.

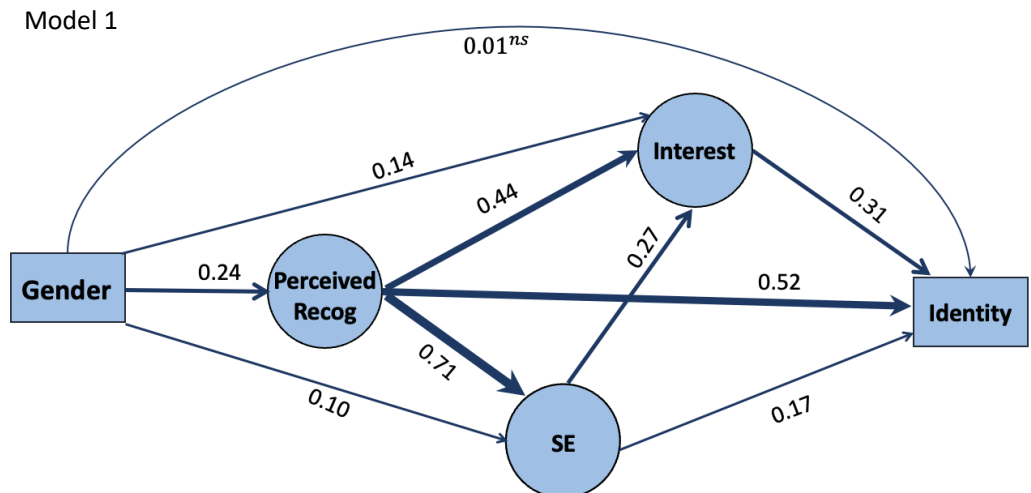

FIG. 2. Results of the path analysis part of SEM, in which perceived recognition predicts interest and self-efficacy, and self-efficacy predicts interest. Each regression line thickness qualitatively corresponds to the magnitude of $\beta$ values. All $\beta$ values shown are significant with $p < 0.001$.

Table V shows the direct and indirect predictive relationships in Model 1. As shown in Table V, even though gender does not directly predict physics identity, it indirectly predicts physics identity with $\beta = 0.28$ through perceived recognition, self-efficacy, and interest, which is consistent with the gender difference shown in Table IV. In addition, we note that perceived recognition not only directly predicts physics identity ($\beta = 0.52$) but also indirectly predicts physics identity ($\beta = 0.32$) through self-efficacy and interest. Moreover, perceived recognition also indirectly predicts interest ($\beta = 0.19$) through self-efficacy. Overall, perceived recognition exhibits strong predictive power for self-efficacy, interest, and physics identity, as indicated by the total regression coefficients. We note that even though self-efficacy indirectly predicts physics identity through interest, this indirect effect is small ($\beta = 0.08$).

**Table V**. Results of the path analysis part of SEM. B represents unstandardized estimate, $\beta$ represent standardized estimate, and SE represents standardized error. The direct effect of gender on physics identity is not statistically significant. All the other non-zero B and $\beta$ values shown are significant with $p < 0.001$.

| Predictor | Outcome | Direct B(SE) | Direct $\beta$ | Indirect B(SE) | Indirect $\beta$ | Total $\beta$ |
|---|---|---|---|---|---|---|
| Gender | Perceived Recognition | 0.51(0.06) | 0.24 | 0.00 | 0.00 | 0.24 |
|  | Self-efficacy | 0.18(0.05) | 0.10 | 0.30(0.04) | 0.17 | 0.27 |
|  | Interest | 0.21(0.04) | 0.14 | 0.26(0.03) | 0.18 | 0.32 |
|  | Physics Identity | 0.03(0.04) | 0.01 | 0.58(0.06) | 0.27 | 0.28 |
| Perceived Recognition | Self-efficacy | 0.58(0.02) | 0.71 | 0.00 | 0.00 | 0.71 |
|  | Interest | 0.31(0.03) | 0.44 | 0.14(0.02) | 0.19 | 0.63 |
|  | Physics Identity | 0.53(0.03) | 0.52 | 0.32(0.02) | 0.32 | 0.84 |
| Self-efficacy | Interest | 0.23(0.04) | 0.27 | 0.00 | 0.00 | 0.27 |
|  | Physics Identity | 0.21(0.04) | 0.17 | 0.10(0.02) | 0.08 | 0.25 |
| Interest | Physics Identity | 0.45(0.04) | 0.31 | 0.00 | 0.00 | 0.31 |

### C. Equivalent SEM Models

The SEM model we specified demonstrates a strong fit with the data. However, as mentioned earlier, a well-fitting model alone is not sufficient to make causal inferences, as there are statistically equivalent models with distinct causal structures that equally well fit the data. Thus, it is crucial to consider these statistically equivalent models, which offer alternative representations of the data, and to evaluate whether one model is more causally accurate compared to the others. Pertaining to RQ3, we found that there are 27 statistically equivalent models in which gender indirectly predicts physics identity through self-efficacy, interest, and perceived recognition while considering the diverse associations among these constructs. There are three possible associations between each pair. These associations are covariance, direct effect via regression from one to the other, or direct effect via regression in the reverse direction. For example, there can be a direct regression path from self-efficacy to interest or from interest to self-efficacy, or there may only be a covariance between self-efficacy and interest. Similarly, there are three possible types of associations between self-efficacy and perceived recognition, and between interest and perceived recognition. Thus, with the constraints that no regression arrows point to gender and arrows can only point to physics identity since it is the outcome variable, there are $3 \times 3 \times 3 = 27$ statistically equivalent SEM models in total. All 27 models have the same fit indices as the model we specified: CFI = 0.998 (>0.90), TLI = 0.998 (>0.90), RMSEA = 0.060 (<0.08) and SRMR = 0.044 (<0.08) [93]. Thus, these statistically equivalent SEM models are all robust from a statistical point of view.

### D. Discussion of different statistically equivalent models

Although all 27 models fit the data equally well, they have different causal structures. Here, we discuss the model we specified, inspired by our prior interviews to make learning environment equitable and inclusive as well as other prior research, along with two other statistically equivalent models. In Model 2, self-efficacy predicts interest and perceived recognition, and interest predicts perceived recognition. In Model 3, interest predicts self-efficacy and perceived recognition, and self-efficacy predicts perceived recognition. We focus on discussing these two equivalent models as they are representative of the models in prior research [3,33,34] and differ from the model we specified by having self-efficacy and interest predicting perceived recognition. Furthermore, we want each of the constructs (perceived recognition, self-efficacy, and interest) to serve as the predictor of the other two once, which can help better illustrate the differences in the causal structures of statistically equivalent models.

As mentioned earlier, Model 1 is aligned with our previous interview findings, which show that women were less likely than men to feel positively recognized by physics instructors/TAs, and this lack of recognition negatively impacted their self-efficacy and interest [22,37-39]. Models 2 and 3 may potentially encourage instructors to reduce the gender gaps in self-efficacy and interest, considering prior research shows that students' interest and self-efficacy are not fixed and that instructors have the ability to increase students' interest in science, technology, engineering, and mathematics (STEM) [71]. However, given that interest-based and self-efficacy-based accounts of gender differences are historically interpreted as fixed [94-96], Models 2 and 3 also have the potential to reinforce college physics instructors' fixed mindset and result in lack of action by them to make more effort to recognize students (particularly those from traditionally marginalized groups such as women due to stereotypes about who can excel in physics) appropriately and create a learning environment with a focus on closing demographic gaps [97].

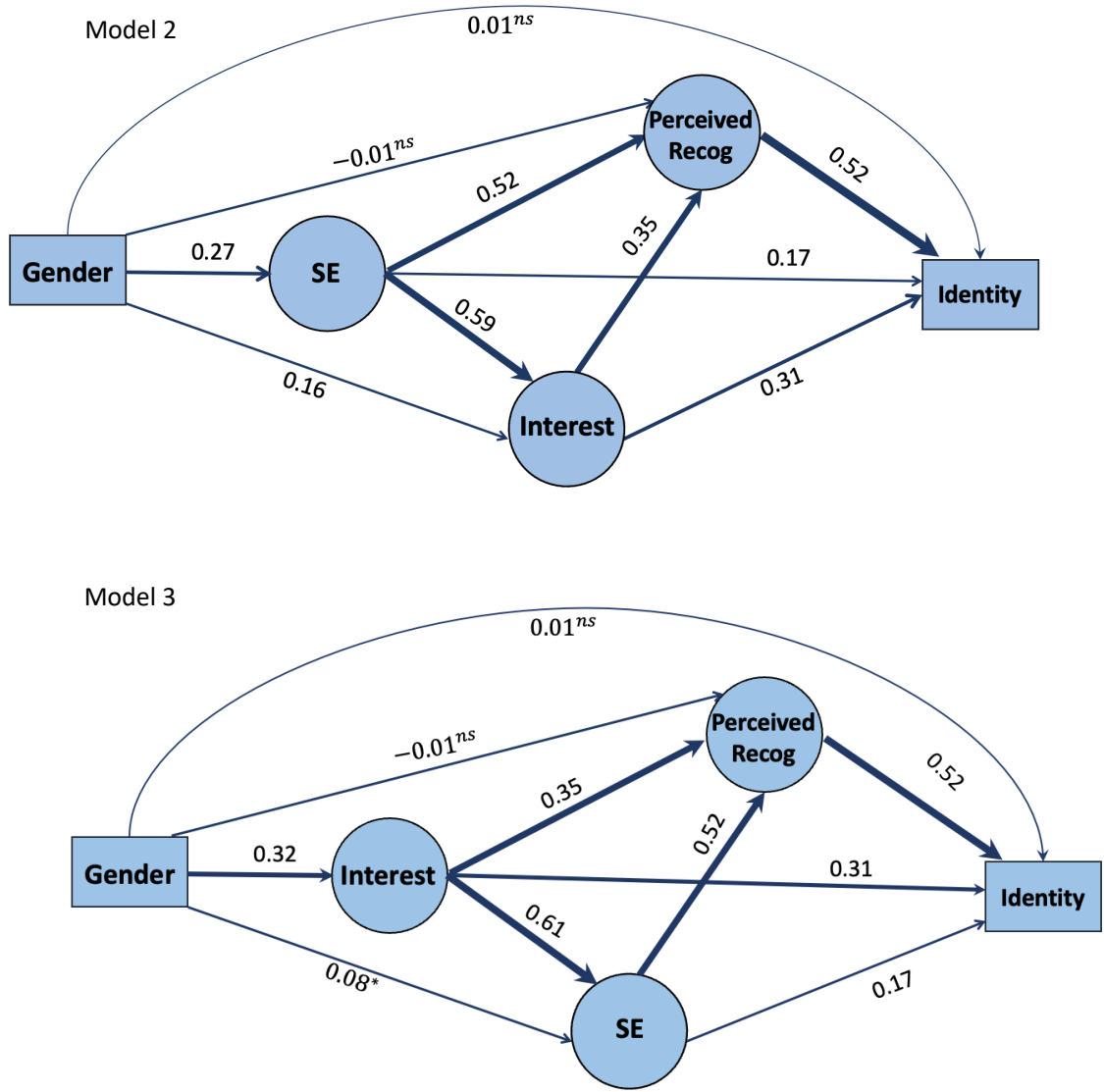

FIG. 3. Results of the path analysis part of the SEM models that show how the relationship between gender and physics identity is mediated through self-efficacy (SE), interest, and perceived recognition (Recog). (a) In Model 2, self-efficacy predicts interest and perceived recognition, and interest predicts perceived recognition. (b) In Model 3, interest predicts self-efficacy and perceived recognition, and self-efficacy predicts perceived recognition. Each regression line thickness qualitatively corresponds to the magnitude of $\beta$ with $0.001 \leqslant p < 0.1$ indicated by ** and $p > 0.05$ indicated by ns. All the other regression lines show relations with $p < 0.001$.

### E. Potential experimental studies to find more accurate causal model from statistically equivalent models

As discussed in earlier sections, we used interview data and other prior studies to specify our model, and in our studies, we have refined our models [20,21] after analyzing greater amounts of interview data [36]. We acknowledge that future studies should continue to gather and consider other additional evidence to further guide and determine a more causally accurate SEM model. Here, we discuss several experimental studies that could help researchers to find more accurate causal model from statistically equivalent models of physics identity.

Previous research suggests that when one set of experiments examines the effect of the independent variable on the mediator, and another set examines the effect of the mediator on the outcome variable, combining these experiments through meta-analysis can provide robust evidence for mediation [13]. Furthermore, if these experiments were conducted in the field, both internal validity and external validity would be maximized [13]. Therefore, we discuss several experimental studies, each aimed at examining specific paths between the variables studied. The paths from perceived recognition to self-efficacy and interest could be investigated by implementing interventions targeting perceived recognition. Similarly, the path from self-efficacy to interest could be investigated by conducting interventions focusing on self-efficacy.

### 1. *Experiments on perceived recognition*

To examine the causal effect of perceived recognition on self-efficacy and interest, one can implement interventions targeting perceived recognition and then observe whether improvements occur in self-efficacy and interest. To establish a causal relationship between perceived recognition and self-efficacy/interest, it is essential to ensure that the intervention solely targets perceived recognition, and the effects of the intervention on self-efficacy and interest should be mediated through perceived recognition.

One potential intervention aimed at enhancing students' perceived recognition involves instructors directly acknowledging their students. For instance, a study demonstrated that students' self-efficacy improves after instructors provide handwritten encouraging feedback on their assignments [44]. This feedback was carefully tailored to each student's performance, incorporating phrases such as "You demonstrated a solid grasp of the concepts!" and "Your solution reflects excellent imagination!" Additionally, feedback included motivational statements like "I note that you are working hard. You have done fine up until now—keep trying." Similarly, a prior study by Mueller and Dweck [98] showed that children who were praised for effort (e.g., "You must have worked hard at these problems.") after an initial task exhibited greater task persistence, greater task enjoyment, and better task performance in subsequent tasks compared to the children who were praised for their intelligence (i.e., "You must be smart at these problems.") or who were simply told they had scored high. These intervention strategies can be implemented by instructors to explicitly recognize their students' performance on quizzes, homework, and exams in one class, which constitutes the treatment group. A suitable control group could be another class taught by the same instructor during the same semester, but this group would not receive the explicit encouraging feedback; instead, they would only receive scores on their homework assignments. Alternatively, the control group could comprise the same instructor's previous students if an instructor does not teach two sections of the same course. To ensure its impact across various instructors, this intervention can be conducted in several different instructors' classes, allowing for a broader assessment of its impact beyond the confines of a single instructor's teaching style.

Another form of perceived recognition intervention draws inspiration from a previous study conducted by Wang and Hazari [99]. This study demonstrates that workshops focusing on training instructors in positively recognizing strategies can effectively support the maintenance or cultivation of students' physics identity development. These strategies encompass acknowledging students' accomplishments and potential [100], encouraging student-led exploration and discussions [101], setting high expectations for students [102], and incorporating activities that foster recognition, such as having students teach a physics concept to a family member or creating physics-related videos for sharing [103]. The workshop conducted by Wang and Hazari revealed that the implementation of these recognition strategies by instructors correlates with a positive shift in students' perception of recognition. This shift, in turn, contributed to the fostering of a stronger physics identity. This workshop can be adapted to act as an intervention for students' perceived recognition by improving instructors' recognition strategies. Researchers can track the implementation of these strategies by instructors and compare students' perceived recognition, self-efficacy, interest, and identity, with those of students taught by the same instructors in previous years before the workshop serving as the control group. Similarly, the workshop should be conducted with multiple instructors to test its effects across different teaching contexts, helping to account for potential instructor-level effects.

For either of the aforementioned perceived recognition interventions, researchers can employ regression analysis or SEM models along with descriptive statistics to examine the effects of the interventions. The pre and post surveys can be conducted with students in the intervention and control classes asking students about the different variables in Fig. 1 to understand, e.g., the impact of perceived recognition on self-efficacy and interest. Researchers can investigate the effect of the intervention on students' post-perceived recognition (measured at the end of the course) after controlling for their pre-perceived recognition, pre-self-efficacy, and pre-interest (measured at the beginning of the course). Similarly, researchers can also explore the intervention's effect on students' post-self-efficacy and post-interest. If, in addition to a positive effect on students' perceived recognition, there is a positive effect on their self-efficacy/interest, then researchers can proceed to conduct SEM to explore whether the effect of the intervention on self-efficacy/interest is mediated through perceived recognition. If the analysis confirms this mediation effect, it can provide evidence to support the causal inference between perceived recognition and self-efficacy/interest as suggested by Model 1. On the other hand, if intervening on perceived recognition causes changes in physics identity with no changes to self-efficacy and interest, then this falsifies Model 1, while Models 2 and 3 remain possible.

In addition, the benefit of the quantitative model is that it proposes quantitative effects (Table V) to be tested against experiment. Although there are many reasons to believe that the quantitative outcomes of intervention will not precisely match observed coefficients, even if a proposed model is the most causally accurate one, observed coefficients provide one benchmark for comparison. For instance, Model 1 predicts that a + 1 standard deviation (SD) increase in perceived recognition should yield a + 0.71 SD increase in self-efficacy, a + 0.63 SD increase in interest, and a + 0.84 SD increase in physics identity. Collecting data on the effects of intervening in perceived recognition and comparing the results against the predictions made by Model 1 can potentially show support for the model or refute it.

## 2. Experiments on self-efficacy

To test the causal effect of self-efficacy on interest, one can conduct interventions targeting students' self-efficacy and observe whether their interest also improves compared to the control group. To establish this causal connection, it is important to craft an intervention centered on self-efficacy, with the effects on interest mediated through self-efficacy.

According to Bandura's social cognitive theory, self-efficacy draws from several sources, including vicarious experience, social persuasion, mastery experience, and psychological state [24]. Vicarious experience, which involves observing others' accomplishments, fosters the belief that similar achievements are attainable. Social persuasion, achieved through encouragement and positive feedback from others, can enhance self-efficacy by affirming an individual's capabilities. Mastery experience, derived from personal achievements and previous successful endeavors, directly boosts self-efficacy. Additionally, an individual's psychological state, encompassing mood and emotions, can significantly influence their self-efficacy. Given that social persuasion closely relates to recognition from others, the self-efficacy intervention discussed here could center on the remaining three sources of self-efficacy.

Many self-efficacy interventions have been developed to enhance students' mathematical self-efficacy, which can be adapted to develop physics self-efficacy intervention. For example, in one study that targeted promoting student mastery experience [104], participants in treatment group were provided with opportunities to successfully complete math-related tasks and receive high score on their performance. The results showed that this intervention had a positive effect on participants' math self-efficacy. In addition, in another study [105] focusing on vicarious experience, a former successful student of a statistics course gave a presentation to the current students about her own math anxieties and shared the strategies that led to her success in the same course. The presentation provided current students with a peer model who had successfully completed the course and resulted in positive impact on student self-efficacy.

To examine the causal relationship from self-efficacy to interest, one can employ similar regression and SEM models as discussed earlier. This approach can help determine whether the self-efficacy intervention not only enhances students' self-efficacy but also positively impacts their interest, and whether the effect on interest is mediated through self-efficacy. If intervening on self-efficacy causes changes in students' interest in physics, this would support the regression path from self-efficacy to interest in Models 1 and 2. In particular, Model 1 predicts that a +1 SD increase in self-efficacy should yield a + 0.27 SD increase in interest and a + 0.25 SD increase in physics identity, while Model 2 predicts that a +1 SD increase in self-efficacy should yield a + 0.59 SD increase in interest and a + 0.73 SD increase in physics identity. These observed coefficients could be used for comparison with the quantitative outcomes of self-efficacy interventions. In addition, it is valuable to examine whether the self-efficacy intervention also leads to improvements in students' perceived recognition and whether self-efficacy predicts recognition. If such an effect is observed, it would support the regression path from self-efficacy to perceived recognition in Models 2 and 3 and suggest that Model 1 is incorrect, or at least incomplete.

### 3. *Experiments on interest*

As discussed earlier, the model we specified is inspired by our prior interviews with students to make physics learning environment equitable and inclusive [36,37] as well other supporting evidence from prior studies [24,32]. For example, our interviews show that a lack of positive perceived recognition can negatively impact students' self-efficacy and interest, and prior studies show that self-efficacy plays an important role in shaping individual's interest. Therefore, the model we specified does not include, e.g., how interest predicts self-efficacy and perceived recognition. However, researchers can implement interventions targeting interest to explore these potential causal relationships, which are suggested by Model 3. For example, prior studies show that effective evidence-based instructional conditions or learning environments that include group work, puzzles, computers etc. can trigger students' situational interest [106-111]. In addition, another study indicated that interventions targeting curricular modifications aligned with the specific interests and experiences of girls as well as enhancing teachers' ability to support girls in developing a positive physics-related self-concept can enhance girls' interest in physics [31]. Researchers can potentially adapt these methods to intervene in college students' interests in physics. Then, by comparing the intervention and control groups, they can evaluate the impact of these interventions on self-efficacy and perceived recognition and whether these effects are mediated by the changes in interest. Model 3 predicts that a + 1 standard deviation (SD) increase in interest should yield a + 0.61 SD increase in self-efficacy and a + 0.67 SD increase in perceived recognition. If targeted interventions on interest also improve students' self-efficacy and perceived recognition, this would support the causal structure of Model 3 and suggest that Model 1 was not causally accurate.

## VII. SUMMARY AND DISCUSSION

In this study, we investigated the predictive relationships among the three dimensions of physics identity: perceived recognition, self-efficacy, and interest. Our results revealed that women scored significantly lower than men in all four motivational constructs, and the gender difference in physics identity is mediated through the gender differences in the other three motivational constructs. Inspired our prior interviews with students to make physics learning environments equitable and inclusive [36,37] and other supporting evidence form prior studies [24,32], we specified a SEM model to describe the predictive relationships among the constructs studied. The statistical analysis shows that the model we specified fits the data well. However, a well-fitting model alone is not sufficient to verify the causal inferences underlying the model, as there are statistically equivalent models with distinct causal structures that equally fit the data. For instance, our model with perceived recognition predicting self-efficacy and interest, emphasizes the role played by instructor recognition. On the other hand, other models with self-efficacy and interest predicting perceived recognition emphasize the significance of students' self-efficacy and interest [112,113]. While our model specification was based on prior interviews with students to make physics learning environments equitable and inclusive [36,37] and other supporting evidence form prior studies [24,32], to find more causally accurate model from among these statistically equivalent models, we discuss several experimental studies. These intervention studies could help test the hypothesized causal effects in different equivalent models. For example, by intervening in one construct, such as perceived recognition, we can evaluate whether students' self-efficacy and interest also improve, thereby further examining the causal relationship between perceived recognition and self-efficacy/interest.

In summary, this paper discusses that a good model fit alone is not sufficient to verify the causal structure of a SEM model due to the existence of statistically equivalent models. Therefore, it is important for researchers to consider statistically equivalent models to the specified model and provide additional evidence for why the proposed model is more accurate than the equivalent ones. In this paper, we discussed several experimental studies that could provide evidence for causal inferences. In this paper, we presented results based on data from the first course of an introductory physics series (Physics 1); see [114] for corresponding data from the second course (Physics 2).

## VIII. LIMITATIONS AND FUTURE DIRECTIONS

In this study, since the gender data were collected by the university using only binary categories, we did not have the gender information of students who did not identify as men or women. This issue has been resolved recently by modifying the way the university is now collecting data. However, since the sample size of these students is small (less than 1% of participants), we would not be able to analyze them as separate groups using multigroup analysis using SEM even if we knew their gender identity beyond women and men. Future studies can use, e.g., qualitative research methods to investigate the motivational beliefs of students in other gender categories. In future studies, we also intend to investigate motivational characteristics of students from other underrepresented groups such as ethnic/racial minority students.

This study examined an introductory calculus-based physics course. It would be valuable to investigate the relationship among women and men's motivational characteristics in other physics courses, e.g., for physics majors. Similar studies in different types of institutions and in other countries would also be helpful for developing a deeper understanding of the relationships among students' motivational characteristics in different contexts.

## ACKNOWLEDGMENTS

This work was supported by Grant No. DUE-1524575 from the National Science Foundation. We would like to thank all students whose data were analyzed and Dr. Robert P. Devaty for his constructive feedback on the manuscript.

# APPENDIX A: CORRELATION BETWEEN ALL ITEMS STUDIED

In the main text, we presented the correlation coefficients between all constructs studied. Here, we present the correlations between all measured indicators as a reference for readers who are interested.

**Table VI**. Correlation coefficients of the items studied using DWLS estimator, along with the mean value (M) and standard deviation (SD) for each item. All correlation coefficients shown are statistically significant with $p < 0.001$. The sample size is N=1203.

| Items | SE1 | SE2 | SE3 | SE4 | Interest 1 | Interest 2 | Interest 3 | Interest 4 | Recog 1 | Recog 2 | Recog 3 | Identity |
|---|---|---|---|---|---|---|---|---|---|---|---|---|
| SE1 | -- | -- | -- | -- | -- | -- | -- | -- | -- | -- | -- | -- |
| SE2 | 0.65 | -- | -- | -- | -- | -- | -- | -- | -- | -- | -- | -- |
| SE3 | 0.58 | 0.66 | -- | -- | -- | -- | -- | -- | -- | -- | -- | -- |
| SE4 | 0.55 | 0.57 | 0.69 | -- | -- | -- | -- | -- | -- | -- | -- | -- |
| Interest 1 | 0.30 | 0.36 | 0.29 | 0.30 | -- | -- | -- | -- | -- | -- | -- | -- |
| Interest 2 | 0.46 | 0.56 | 0.47 | 0.40 | 0.65 | -- | -- | -- | -- | -- | -- | -- |
| Interest 3 | 0.43 | 0.49 | 0.39 | 0.41 | 0.64 | 0.73 | -- | -- | -- | -- | -- | -- |
| Interest 4 | 0.36 | 0.43 | 0.34 | 0.34 | 0.54 | 0.65 | 0.66 | -- | -- | -- | -- | -- |
| Recog 1 | 0.56 | 0.56 | 0.47 | 0.45 | 0.48 | 0.60 | 0.57 | 0.48 | -- | -- | -- | -- |
| Recog 2 | 0.62 | 0.55 | 0.50 | 0.50 | 0.46 | 0.61 | 0.55 | 0.47 | 0.88 | -- | -- | -- |
| Recog 3 | 0.58 | 0.57 | 0.55 | 0.49 | 0.34 | 0.47 | 0.45 | 0.37 | 0.69 | 0.69 | -- | -- |
| Identity | 0.60 | 0.62 | 0.58 | 0.53 | 0.51 | 0.69 | 0.65 | 0.54 | 0.83 | 0.82 | 0.70 | -- |
| Mean | 2.73 | 2.97 | 2.97 | 2.88 | 3.12 | 3.03 | 2.82 | 2.89 | 2.57 | 2.55 | 2.30 | 2.47 |
| SD | 0.71 | 0.60 | 0.75 | 0.66 | 0.84 | 0.71 | 0.79 | 0.77 | 0.88 | 0.86 | 0.78 | 0.86 |

# APPENDIX B: PERCENTAGES OF STUDENTS WHO SELECTED EACH CHOICE FOR EACH SURVEY ITEM

In the main text, we discussed an investigation of women's and men's self-efficacy, interest, perceived recognition and physics identity. Here, we present the percentages of women and men who selected each answer choice from a 4-point Likert scale for each survey item. Students were given a score from 1 to 4 respectively with higher scores indicating greater levels of self-efficacy, interest, perceived recognition, and physics identity.

By comparing percentages of women and men who selected each answer choice, we found that for all survey items, the percentages of women who selected 1 or 2 were larger than those of men, while the percentages of women who selected 4 were smaller than those of men. These findings are consistent with Table IV showing that there were statistically significant gender differences in all motivational constructs studied.

**TABLE VII.** Percentages of women and men who selected each choice from a 4-point Likert scale for each survey item of self-efficacy (SE) in the pre- and post-survey, which have the response scale: 1= NO!, 2 = no, 3 = yes, and 4 = YES!. Mean represents the mean score value of the item for women and men separately, SD represents the standard deviation of this item.

| Gender | Survey items | 1 | 2 | 3 | 4 | Mean | SD |
|---|---|---|---|---|---|---|---|
| **Women** | SE1 | 10% | 31% | 53% | 6% | 2.56 | 0.76 |
|  | SE2 | 4% | 18% | 70% | 8% | 2.81 | 0.63 |
|  | SE3 | 6% | 29% | 52% | 13% | 2.73 | 0.76 |
|  | SE4 | 5% | 25% | 60% | 10% | 2.73 | 0.70 |
| **Men** | SE1 | 4% | 22% | 63% | 11% | 2.82 | 0.67 |
|  | SE2 | 1% | 10% | 70% | 19% | 3.06 | 0.57 |
|  | SE3 | 2% | 14% | 56% | 28% | 3.09 | 0.71 |
|  | SE4 | 1% | 17% | 66% | 16% | 2.96 | 0.62 |

**TABLE VIII.** Percentages of women and men who selected each choice from a 4-point Likert scale for each survey item of interest in the pre- and post-survey. Interest1 has the response scale: 1 = Never, 2 = Once a month, 3 = Once a week, 4 = Every day". Interest2 has the response scale: 1 = Very boring, 2 = boring, 3 = interesting, 4 = Very interesting. The other two items have the response scale: 1= NO!, 2 = no, 3 = yes, and 4 = YES!. Mean represents the mean score value of the item for women and men separately, SD represents the standard deviation of this item.

| Gender | Survey items | 1 | 2 | 3 | 4 | Mean | SD |
|---|---|---|---|---|---|---|---|
| Women | Interest 1 | 8% | 21% | 47% | 24% | 2.85 | 0.88 |
|  | Interest 2 | 7% | 19% | 61% | 13% | 2.81 | 0.74 |
|  | Interest 3 | 7% | 42% | 41% | 10% | 2.54 | 0.76 |
|  | Interest 4 | 7% | 31% | 50% | 12% | 2.66 | 0.78 |
| Men | Interest 1 | 3% | 11% | 42% | 44% | 3.26 | 0.79 |
|  | Interest 2 | 2% | 9% | 61% | 28% | 3.15 | 0.66 |
|  | Interest 3 | 3% | 22% | 50% | 25% | 2.98 | 0.76 |
|  | Interest 4 | 3% | 19% | 53% | 25% | 3.01 | 0.74 |

**TABLE IX.** Percentages of women and men who selected each choice from a 4-point Likert scale for each survey item of perceived recognition and physics identity. All items have the response scale: 1 = strongly disagree, 2 = disagree, 3 = agree, and 4 = strongly agree. Mean represents the mean score value of the item for women and men separately, SD represents the standard deviation of this item.

| Gender | Survey items | 1 | 2 | 3 | 4 | Mean | SD |
|---|---|---|---|---|---|---|---|
| Women | Recognition 1 | 19% | 40% | 34% | 7% | 2.30 | 0.86 |
|  | Recognition 2 | 17% | 40% | 35% | 7% | 2.33 | 0.85 |
|  | Recognition 3 | 22% | 47% | 29% | 2% | 2.12 | 0.76 |
|  | Identity | 22% | 45% | 27% | 6% | 2.17 | 0.83 |
| Men | Recognition 1 | 8% | 29% | 45% | 18% | 2.72 | 0.85 |
|  | Recognition 2 | 9% | 32% | 43% | 16% | 2.67 | 0.85 |
|  | Recognition 3 | 11% | 42% | 41% | 6% | 2.41 | 0.77 |
|  | Identity | 8% | 35% | 43% | 14% | 2.63 | 0.83 |

# APPENDIX C: LOCAL FIT OF THE SEM MODEL

In the main text, we reported the fit indices showing that our model fit the data well. In addition to global fit, local fit can provide a deeper insight into the extent to which our model aligns with the data. Here, we report the results of two measures of local fit. Table XI illustrates the residual correlations among the studied items. The results reveal that all residual correlations between the items are notably small, indicating that our model effectively accounts for most correlations among these items. In addition, we examine the modification indices of our model. Modification index larger than 3.84 indicates that the model fit would be significantly improved, and the $p$ value for the added parameter would be < 0.05 [115,116]. In our model, the only path omitted is from gender to physics identity, and the modification index associated with this path is 0.128—significantly lower than the threshold of 3.84. Therefore, our model's local fit is also good.

**TABLE X.** Residual correlation between the items studied. SE represents self-efficacy, and Recog represents perceived recognition.

| Items | SE1 | SE2 | SE3 | SE4 | Interest 1 | Interest 2 | Interest 3 | Interest 4 | Recog 1 | Recog 2 | Recog 3 | Identity |
|---|---|---|---|---|---|---|---|---|---|---|---|---|
| SE1 | -- | -- | -- | -- | -- | -- | -- | -- | -- | -- | -- | -- |
| SE2 | -0.01 | -- | -- | -- | -- | -- | -- | -- | -- | -- | -- | -- |
| SE3 | -0.06 | 0.00 | -- | -- | -- | -- | -- | -- | -- | -- | -- | -- |
| SE4 | -0.05 | -0.04 | 0.11 | -- | -- | -- | -- | -- | -- | -- | -- | -- |
| Interest 1 | -0.06 | 0.01 | -0.08 | -0.03 | -- | -- | -- | -- | -- | -- | -- | -- |
| Interest 2 | 0.01 | 0.10 | 0.02 | -0.02 | 0.01 | -- | -- | -- | -- | -- | -- | -- |
| Interest 3 | 0.00 | 0.03 | -0.05 | 0.01 | 0.04 | -0.05 | -- | -- | -- | -- | -- | -- |
| Interest 4 | -0.02 | 0.04 | -0.03 | -0.01 | 0.02 | -0.02 | 0.03 | -- | -- | -- | -- | -- |
| Recog 1 | -0.03 | -0.05 | -0.11 | -0.09 | 0.02 | 0.00 | 0.00 | -0.01 | -- | -- | -- | -- |
| Recog 2 | 0.04 | -0.05 | -0.07 | -0.04 | 0.00 | 0.02 | 0.00 | -0.01 | 0.01 | -- | -- | -- |
| Recog 3 | 0.13 | 0.10 | 0.11 | 0.07 | -0.03 | 0.01 | -0.01 | -0.01 | -0.08 | -0.08 | -- | -- |
| Identity | 0.01 | 0.00 | 0.00 | -0.02 | -0.02 | 0.01 | 0.01 | -0.03 | 0.00 | -0.02 | 0.05 | -- |

[1] R. B. Kline, *Principles and Practice of Structural Equation Modeling* (Guilford publications, 2015).
[2] Y. Li and C. Singh, Effect of gender, self-efficacy, and interest on perception of the learning environment and outcomes in calculus-based introductory physics courses, Phys. Rev. Phys. Educ. Res. **17**, 010143 (2021).
[3] A. Godwin, G. Potvin, Z. Hazari, and R. Lock, Identity, critical agency, and engineering: An affective model for predicting engineering as a career choice, J. Engin. Educ. **105**, 312 (2016).
[4] V. Adlakha and E. Kuo, Critical issues in statistical causal inference for observational physics education research, Phys. Rev. Phys. Educ. Res. **19**, 020160 (2023).
[5] G. Shmueli, To explain or to predict?, (2010).
[6] M. Kuhn and K. Johnson, *Applied predictive modeling* (Springer, 2013), Vol. 26.
[7] J. Pearl and D. Mackenzie, *The book of why: the new science of cause and effect* (Basic books, 2018).
[8] R. C. MacCallum and J. T. Austin, Applications of structural equation modeling in psychological research, Annual review of psychology **51**, 201 (2000).
[9] D. Kaplan, *Structural equation modeling: Foundations and extensions* (Sage Publications, 2008), Vol. 10.
[10] S. L. Hershberger and G. Marcoulides, The problem of equivalent structural models, in *Structural equation modeling: A second course*, edited by G. R. Hancock and R. O. Mueller (IAP, 2006), p. 13.
[11] D. B. Rubin, Comment: The design and analysis of gold standard randomized experiments, Journal of the American Statistical Association **103**, 1350 (2008).
[12] S. A. Mulaik, *Linear causal modeling with structural equations* (CRC press, 2009).
[13] D. Eden, E. F. Stone-Romero, and H. R. Rothstein, Synthesizing results of multiple randomized experiments to establish causality in mediation testing, Human Resource Management Review **25**, 342 (2015).
[14] J. G. Bullock, D. P. Green, and S. E. Ha, Yes, but what's the mechanism?(don't expect an easy answer), J. Pers. Soc. Psychol. **98**, 550 (2010).
[15] S. J. Spencer, M. P. Zanna, and G. T. Fong, Establishing a causal chain: why experiments are often more effective than mediational analyses in examining psychological processes, J. Pers. Soc. Psychol. **89**, 845 (2005).
[16] Y. Kifer, D. Heller, W. Q. E. Perunovic, and A. D. Galinsky, The good life of the powerful: The experience of power and authenticity enhances subjective well-being, Psychol. Sci. **24**, 280 (2013).
[17] Z. Hazari, G. Sonnert, P. M. Sadler, and M.-C. Shanahan, Connecting high school physics experiences, outcome expectations, physics identity, and physics career choice: A gender study, J. Res. Sci. Teach. **47**, 978 (2010).
[18] Y. Li and C. Singh, Inclusive learning environments can improve student learning and motivational beliefs, Phys. Rev. Phys. Educ. Res. **18**, 020147 (2022).
[19] Z. Hazari and C. Cass, Towards meaningful physics recognition: What does this recognition actually look like?, The Phys. Teach. **56**, 442 (2018).
[20] Z. Y. Kalender, E. Marshman, C. D. Schunn, T. J. Nokes-Malach, and C. Singh, Why female science, technology, engineering, and mathematics majors do not identify with physics: They do not think others see them that way, Phys. Rev. Phys. Educ. Res. **15**, 020148 (2019).

[21] Y. Li and C. Singh, Do female and male students' physics motivational beliefs change in a two-semester introductory physics course sequence?, Phys. Rev. Phys. Educ. Res. **18**, 010142 (2022).
[22] Y. Li, K. Whitcomb, and C. Singh, How perception of being recognized or not recognized by instructors as a "physics person" impacts male and female students' self-efficacy and performance, The Phys. Teach. **58**, 484 (2020).
[23] S. Cwik and C. Singh, Not feeling recognized as a physics person by instructors and teaching assistants is correlated with female students' lower grades, Phys. Rev. Phys. Educ. Res. **18**, 010138 (2022).
[24] A. Bandura, Self-efficacy, in *Encyclopedia of Psychology*, 2nd ed., edited by R. J. Corsini (Wiley, New York, 1994), Vol. 3, p. 368.
[25] P. Vincent-Ruz and C. D. Schunn, The increasingly important role of science competency beliefs for science learning in girls, J. Res. Sci. Teach. **54**, 790 (2017).
[26] B. J. Zimmerman, Self-efficacy: An essential motive to learn, Contemp. Educ. Psychol. **25**, 82 (2000).
[27] D. H. Schunk and F. Pajares, The Development of Academic Self-Efficacy, in *Development of Achievement Motivation: A Volume in the Educational Psychology Series*, edited by A. Wigfield and J. S. Eccles (Academic Press, San Diego, 2002), p. 15.
[28] H. M. Watt, The role of motivation in gendered educational and occupational trajectories related to maths, Educ. Res. Eval. **12**, 305 (2006).
[29] S. Hidi, Interest: A unique motivational variable, Educ. Res. Rev. **1**, 69 (2006).
[30] J. L. Smith, C. Sansone, and P. H. White, The stereotyped task engagement process: The role of interest and achievement motivation, J. Educ. Psychol. **99**, 99 (2007).
[31] P. Häussler and L. Hoffmann, An intervention study to enhance girls' interest, self-concept, and achievement in physics classes, J. Res. Sci. Teach. **39**, 870 (2002).
[32] S. Hidi and K. A. Renninger, The four-phase model of interest development, Educ. Psychol. **41**, 111 (2006).
[33] R. M. Lock, Z. Hazari, and G. Potvin, Impact of out-of-class science and engineering activities on physics identity and career intentions, Phys. Rev. Phys. Educ. Res. **15**, 020137 (2019).
[34] Z. Hazari, D. Chari, G. Potvin, and E. Brewe, The context dependence of physics identity: Examining the role of performance/competence, recognition, interest, and sense of belonging for lower and upper female physics undergraduates, J. Res. Sci. Teach. **57**, 1583 (2020).
[35] S. A. Crockett, A five-step guide to conducting SEM analysis in counseling research, Counseling Outcome Research and Evaluation **3**, 30 (2012).
[36] Y. Li and C. Singh, Impact of perceived recognition by physics instructors on women's self-efficacy and interest, Phys. Rev. Phys. Educ. Res. **19**, 020125 (2023).
[37] D. Doucette and C. Singh, Why are there so few women in physics? Reflections on the experiences of two women, The Phys. Teach. **58**, 297 (2020).
[38] L. M. Santana and C. Singh, Negative impacts of an unwelcoming physics environment on undergraduate women, in *Physics Education Research Conference 2021 Virtual Conference* (2021), p. 377.
[39] D. Doucette, R. Clark, and C. Singh, Hermione and the Secretary: How gendered task division in introductory physics labs can disrupt equitable learning, Eur. J. Phys. **41**, 035702 (2020).

[40] L. Santana and C. Singh, *Investigating experiences of a Black woman in physics and astronomy* 2022).
[41] L. M. Santana and C. Singh, Negative impacts of an unwelcoming physics environment on undergraduate women, in *Proceedings of the Physics Education Research Conference (PERC)* (2021), p. 377.
[42] J. C. Chan and S.-f. Lam, Effects of different evaluative feedback on students' self-efficacy in learning, Instructional Science **38**, 37 (2010).
[43] R. Ruegg, The effect of peer and teacher feedback on changes in EFL students' writing self-efficacy, The Language Learning Journal **46**, 87 (2018).
[44] B. W. Tuckman and T. L. Sexton, The effect of teacher encouragement on student self-efficacy and motivation for self-regulated performance, Journal of Social Behavior and Personality **6**, 137 (1991).
[45] J. Henderlong and M. R. Lepper, The effects of praise on children's intrinsic motivation: a review and synthesis, Psychological bulletin **128**, 774 (2002).
[46] M. E. Shanab, D. Peterson, S. Dargahi, and P. Deroian, The effects of positive and negative verbal feedback on the intrinsic motivation of male and female subjects, The Journal of Social Psychology **115**, 195 (1981).
[47] E. L. Deci, R. Koestner, and R. M. Ryan, A meta-analytic review of experiments examining the effects of extrinsic rewards on intrinsic motivation, Psychological bulletin **125**, 627 (1999).
[48] D. A. Donnay and F. H. Borgen, The incremental validity of vocational self-efficacy: An examination of interest, self-efficacy, and occupation, Journal of Counseling Psychology **46**, 432 (1999).
[49] R. W. Lent, S. D. Brown, and G. Hackett, Toward a unifying social cognitive theory of career and academic interest, choice, and performance, Journal of vocational behavior **45**, 79 (1994).
[50] N. E. Betz, Self-efficacy theory as a basis for career assessment, Journal of career Assessment **8**, 205 (2000).
[51] P. J. Silvia, Interest and interests: The psychology of constructive capriciousness, Review of General Psychology **5**, 270 (2001).
[52] P. J. Silvia, Self-efficacy and interest: Experimental studies of optimal incompetence, Journal of Vocational Behavior **62**, 237 (2003).
[53] D. E. Berlyne, Conflict, arousal, and curiosity,  (1960).
[54] S. Cwik and C. Singh, How perception of learning environment predicts male and female students' grades and motivational outcomes in algebra-based introductory physics courses, Phys. Rev. Phys. Educ. Res. **17**, 020143 (2021).
[55] E. Marshman, Y. Kalender, T. Nokes-Malach, C. Schunn, and C. Singh, Female students with A's have similar physics self-efficacy as male students with C's in introductory courses: A cause for alarm?, Phys. Rev. Phys. Educ. Res. **14**, 020123 (2018).
[56] G. C. Marchand and G. Taasoobshirazi, Stereotype threat and women's performance in physics, Int. J. Sci. Educ. **35**, 3050 (2013).
[57] A. Maries, N. I. Karim, and C. Singh, Is agreeing with a gender stereotype correlated with the performance of female students in introductory physics?, Phys. Rev. Phys. Educ. Res. **14**, 020119 (2018).
[58] A. Maries, N. I. Karim, and C. Singh, Does stereotype threat affect female students' performance in introductory physics?, in *AIP Conf. Proc.*, AIP Publishing LLC (2019), p. 120001.

[59] A. A. Eaton, J. F. Saunders, R. K. Jacobson, and K. West, How Gender and Race Stereotypes Impact the Advancement of Scholars in STEM: Professors' Biased Evaluations of Physics and Biology Post-Doctoral Candidates, Sex roles **82**, 127 (2020).

[60] Y. Li and C. Singh, Sense of belonging is an important predictor of introductory physics students' academic performance, Phys. Rev. Phys. Educ. Res. **19**, 020137 (2023).

[61] Learning Activation Lab, Activation lab tools: Measures and data collection instruments (2017), http://www.activationlab.org/tools/. (Accessed 4 February 2019).

[62] J. Schell and B. Lukoff, Peer instruction self-efficacy instrument [Developed at Harvard University] (unpublished), (2010).

[63] S. M. Glynn, P. Brickman, N. Armstrong, and G. Taasoobshirazi, Science motivation questionnaire II: Validation with science majors and nonscience majors, J. Res. Sci. Teach. **48**, 1159 (2011).

[64] E. Marshman, Z. Y. Kalender, C. Schunn, T. Nokes-Malach, and C. Singh, A longitudinal analysis of students' motivational characteristics in introductory physics courses: Gender differences, Can. J. Phys. **96**, 391 (2018).

[65] T. Nokes-Malach, E. Marshman, Z. Y. Kalender, C. Schunn, and C. Singh, Investigation of male and female students' motivational characteristics throughout an introductory physics course sequence, in *Proceedings of the 2017 Physics Education Research Conference Cincinnati, OH* (2017), p. 276.

[66] Z. Y. Kalender, E. Marshman, C. D. Schunn, T. J. Nokes-Malach, and C. Singh, Large gender differences in physics self-efficacy at equal performance levels: A warning sign?, in *Proceeding of the 2018 Physics Education Research Conference Washington, DC* (2018).

[67] B. Thompson, *Exploratory and Confirmatory Factor Analysis* (American Psychological Association, Washington, 2004).

[68] D. Hooper, J. Coughlan, and M. Mullen, Structural equation modeling: Guidelines for determining model fit, Electronic J. Bus. Res. Methods **6**, 53 (2007).

[69] L. J. Cronbach, Coefficient alpha and the internal structure of tests, Psychometrika **16**, 297 (1951).

[70] K. Pearson and F. Galton, VII. Note on regression and inheritance in the case of two parents, Proc. Royal Soc. London **58**, 240 (1895).

[71] Z. Hazari, G. Potvin, R. M. Lock, F. Lung, G. Sonnert, and P. M. Sadler, Factors that affect the physical science career interest of female students: Testing five common hypotheses, Phys. Rev. ST Phys. Educ. Res. **9**, 020115 (2013).

[72] Z. Hazari, R. H. Tai, and P. M. Sadler, Gender differences in introductory university physics performance: The influence of high school physics preparation and affective factors, Sci. Educ. **91**, 847 (2007).

[73] R. Likert, A technique for the measurement of attitudes, Arch. Sci. Psychol. **22**, 55 (1932).

[74] S. E. Embretson and S. P. Reise, *Item Response Theory for Psychologists* (Lawrence Erlbaum Associates Publishers, Mahwah, NJ, US, 2000).

[75] R. P. Chalmers, mirt: A multidimensional item response theory package for the R environment, Journal of Statistical Software **48**, 1 (2012).

[76] F. Samejima, Estimation of latent ability using a response pattern of graded scores, in *Psychometrika Monograph* (Psychometric Society, Richmond, VA, 1969), p. 17.

[77] M. G. Bulmer, *Principles of statistics* (Courier Corporation, 1979).

[78] J. F. Hair, W. C. Black, B. J. Babin, and R. E. Anderson, *Multivariate Data Analysis: International Version* (Pearson, New Jersey, 2010).

[79] B. M. Byrne, *Structural equation modeling with AMOS: Basic concepts, applications, and programming, 2nd ed* (Routledge/Taylor & Francis Group, New York, NY, US, 2010), Structural equation modeling with AMOS: Basic concepts, applications, and programming, 2nd ed.
[80] J. Pallant, *SPSS survival manual: A step by step guide to data analysis using IBM SPSS* (Routledge, 2020).
[81] R Core Team, R: A Language and Environment for Statistical Computing (2013), (R Foundation) https://www.r-project.org/ (Accessed 6 February 2019).
[82] A. J. Tomarken and N. G. Waller, Structural equation modeling: Strengths, limitations, and misconceptions, Annu. Rev. Clin. Psychol. **1**, 31 (2005).
[83] S. J. Finney and C. DiStefano, Non-normal and categorical data in structural equation modeling, Structural equation modeling: A second course **10**, 269 (2006).
[84] D. Mindrila, Maximum likelihood (ML) and diagonally weighted least squares (DWLS) estimation procedures: A comparison of estimation bias with ordinal and multivariate non-normal data, International Journal of Digital Society **1**, 60 (2010).
[85] M. Rhemtulla, P. É. Brosseau-Liard, and V. Savalei, When can categorical variables be treated as continuous? A comparison of robust continuous and categorical SEM estimation methods under suboptimal conditions, Psychological methods **17**, 354 (2012).
[86] K. S. Betts, G. M. Williams, J. M. Najman, and R. Alati, The role of sleep disturbance in the relationship between post-traumatic stress disorder and suicidal ideation, Journal of anxiety disorders **27**, 735 (2013).
[87] C. A. Kronauge, *The effects of mixing metrics and distributions simultaneously in structural equation modeling: A simulation study* (University of Northern Colorado, 2012).
[88] H. Akoglu, User's guide to correlation coefficients, Turkish Journal of Emergency Medicine **18**, 91 (2018).
[89] G. D. Garson, *Structural equation modeling* (G. David Garson and Statistical Publishing Associates, 2014).
[90] J. Cohen, *Statistical Power Analysis for the Behavioral Sciences* (L. Erlbaum Associates, Hillsdale, N.J., 1988).
[91] J. Cohen, *Statistical Power Analysis for the Behavioral Sciences* (Routledge, 2013).
[92] J. B. Grace and K. A. Bollen, Interpreting the results from multiple regression and structural equation models, Bulletin of the Ecological Society of America **84**, 283 (2005).
[93] D. Hooper, J. Coughlan, and M. R. Mullen, Structural equation modelling: Guidelines for determining model fit, Electronic J. Bus. Res. Methods **6**, pp53-60 (2008).
[94] M. Seitchik, Confidence and gender: Few differences, but gender stereotypes impact perceptions of confidence, The Psychologist-Manager Journal **23**, 194 (2020).
[95] A. Master, A. N. Meltzoff, and S. Cheryan, Gender stereotypes about interests start early and cause gender disparities in computer science and engineering, Proceedings of the National Academy of Sciences **118**, e2100030118 (2021).
[96] J. Wade, How you can change gender stereotypes about physicists, Nature Reviews Physics **4**, 690 (2022).
[97] C. Singh, Inclusive mentoring: The mindset of an effective mentor, APS News **30** (2021).
[98] C. M. Mueller and C. S. Dweck, Praise for intelligence can undermine children's motivation and performance, J. Pers. Soc. Psychol. **75**, 33 (1998).
[99] J. Wang and Z. Hazari, Promoting high school students' physics identity through explicit and implicit recognition, Phys. Rev. Phys. Educ. Res. **14**, 020111 (2018).
[100] A. Bandura, *Self-efficacy: The Exercise of Control* (W. H. Freeman, New York, 1997).

[101] W. Speering and L. Rennie, Students' perceptions about science: The impact of transition from primary to secondary school, Research in Science Education **26**, 283 (1996).

[102] Z. Hazari, C. Cass, and C. Beattie, Obscuring power structures in the physics classroom: Linking teacher positioning, student engagement, and physics identity development, J. Res. Sci. Teach. **52**, 735 (2015).

[103] R. Elmesky, "I am science and the world is mine": Embodied practices as resources for empowerment, School Science and Mathematics **105**, 335 (2005).

[104] D. A. Luzzo, P. Hasper, K. A. Albert, M. A. Bibby, and E. A. Martinelli Jr, Effects of self-efficacy-enhancing interventions on the math/science self-efficacy and career interests, goals, and actions of career undecided college students, Journal of Counseling Psychology **46**, 233 (1999).

[105] R. A. Bartsch, K. A. Case, and H. Meerman, Increasing academic self-efficacy in statistics with a live vicarious experience presentation, Teaching of Psychology **39**, 133 (2012).

[106] D. Cordova and M. Lepper, Intrinsic motivation and the process of learning: Beneficial effects of contextualization, personalization, and choice, J. Educ. Psychol. **88**, 715 (1996).

[107] S. Hidi and W. Baird, Strategies for increasing text-based interest and students' recall of expository texts, Reading Research Quarterly **23**, 465 (1988).

[108] S. Hidi, J. Weiss, D. Berndorff, and J. Nolan, The role of gender, instruction and a cooperative learning technique in science education across formal and informal settings, in *Interest and learning: Proceedings of the Seeon conference on interest and gender*, IPN Kiel, Germany (1998), p. 215.

[109] M. R. Lepper and D. I. Cordova, A desire to be taught: Instructional consequences of intrinsic motivation, Motivation and Emotion **16**, 187 (1992).

[110] M. Mitchell, Situational interest: Its multifaceted structure in the secondary school mathematics classroom, J. Educ. Psychol. **85**, 424 (1993).

[111] J. A. Sloboda and J. Davidson, The young performing musician, in *Musical Beginnings: Origins and Development of Musical Competence* (Oxford University Press, 1996), p. 171.

[112] E. A. Canning, K. Muenks, D. J. Green, and M. C. Murphy, STEM faculty who believe ability is fixed have larger racial achievement gaps and inspire less student motivation in their classes, Sci. Adv. **5**, eaau4734 (2019).

[113] C. S. Dweck, *Mindset: The New Psychology of Success* (Random House Digital, Inc., 2008).

[114] Y. Li and C. Singh, Considering Statistically Equivalent Models when using Structural Equation Modeling: an Example from Physics Identity, International Journal of Innovation in Science and Mathematics Education **31** (2023).

[115] P. W. Lei and Q. Wu, Introduction to structural equation modeling: Issues and practical considerations, Educational Measurement: issues and practice **26**, 33 (2007).

[116] R. P. Bagozzi and Y. Yi, On the evaluation of structural equation models, Journal of the academy of marketing science **16**, 74 (1988).